\documentclass[final,english]{bullsrsl}[2023/06/15]

\RequirePackage{ifthen}
\RequirePackage{mathptmx}
\RequirePackage{hyperref}
\RequirePackage{lineno}
\RequirePackage{geometry}

\usepackage[latin1]{inputenc}
\usepackage[T1]{fontenc}

\usepackage{natbib} 
\usepackage{graphicx}
\usepackage{threeparttable}
\begin{document}
\title{Rotational Variability and Detection of Superflares in a Young Brown Dwarf by TESS}
\author[corresponding]{Rajib}{Kumbhakar}
\author[corresponding]{Soumen}{Mondal}
\author[]{Samrat}{Ghosh}
\author[]{Diya}{Ram}
\author[]{Sudip}{Pramanik}
\affiliation[]{S. N. Bose National Centre for Basic Sciences, Salt Lake, Kolkata 700106, India}
\correspondance{rajskbuph0@gmail.com; soumen.mondal@bose.res.in}
\date{20th August 2023}
\maketitle
%

\begin{abstract}
We present a comprehensive analysis of a Transiting Exoplanet Survey Satellite (TESS) high-quality light curve for a young brown dwarf, MHO~4 having spectral type M7.0, in the Taurus star-forming region. We investigate the rotation periods and characterize the BD's dynamic atmosphere and surface features. We present light curve analysis of MHO~4, and estimate the rotation period to be around 2.224~d. Remarkably, MHO~4 exhibits two significant flaring events.  Furthermore, we also estimated bolometric flare energies to be within the energy range of $10^{34}$ to $10^{35}$ erg, which sits in the superflare category.

\end{abstract}


\section{Introduction}
Brown dwarfs (BDs) were initially theorized by Shiv Kumar in 1962 \citep{Joergens2014ASSL..401....1J} and so far thousands of BDs have been discovered \citep{Martin1999AJ....118.2466M, Carnero2019MNRAS.489.5301C}. BDs occupy an intermediate-mass range between planets and stars, typically ranging from approximately 13 to 80 times the mass of Jupiter. Due to their low masses, they can not sustain nuclear fusion by burning hydrogen inside their core; instead, they burn deuterium. 
Being cool and fully convective, BDs show different kinds of atmospheric features and cause modulation of flux as it rotates, which can be interpreted by studying the light curve. Therefore, photometric variability is an important tool for studying their surface characteristics and atmospheric properties.

In recent years, variability of these objects has been observed over a broad range of wavelengths \citep{Carpenter2001AJ....121.3160C, Ghosh2021MNRAS.500.5106G, Wang2023arXiv230502514W}. In the optical range, photometric variability of BDs mainly arises due to the rotational modulation of active starspots regions in the photosphere, along with magnetically induced chromospheric activity at higher altitudes in the atmosphere, dust clouds, or binary companions. Thus variability can probe the diverse set of several physical mechanisms for these classes of young objects and study diverse environments. With age, the shape of light curves of young objects, their rotation periods, and surface features in the photosphere has been investigated by the availability of high-cadence and high-precision photometric observations from spacecraft such as CoRoT \citep{Baglin2006cosp...36.3749B}, MOST \citep{Matthews2000ASPC..203...74M}, and Kepler extended mission K2 \cite{Howell2014PASP..126..398H}. Here we used the Transiting Exoplanet Survey Satellite (TESS; \citet{Ricker2015JATIS...1a4003R}), which also has high-cadence, high-precision light curves of young objects, covering many young clusters, including the extended Taurus star-forming region with wavelength range from 600 to 1000 $nm$.

In this study, we presented the analysis of TESS photometric data of a young BD, MHO~4 (TIC 456944264) with spectral type M7.0 \citep{White2003ApJ...582.1109W}. It is a bonafide member of the young Taurus star-forming region. We further analyze the power spectrum to infer the rotation periods and the physical mechanism behind stellar flares. 
Section 2 describes the TESS observation and data analysis procedure, including period analysis. In Section 3, we presented the light curve of our source and described flare analysis and the results derived from them. Finally, in Section 4, we summarized the key findings of our study. 

\section{TESS Observations and Data Analysis}
We used the time-series data of TESS, which is briefly discussed here (for more details, please see \citealt{Ricker2015JATIS...1a4003R}). TESS is a space-based telescope launched by NASA in April 2018. It consists of four cameras with a size of 10 cm and a combined field of view $24^{\circ} \times 96^{\circ}$. During the first extended mission, TESS observed our selected object, MHO 4, in both TESS sector 43 (Camera 3 and CCD 3) and sector 44 (Camera 2, CCD 4). We used 2-min cadence photometry, publicly accessible through the Mikulski Archive for Space Telescopes (MAST). We use \texttt{lightkurve} \citep{Lightkurve_Collaboration_2018ascl.soft12013L} package to download the lightcurve of MHO 4 from MAST. These data were already processed by the Science Processing Operations Center (SPOC), and we used Pre-search Data Conditioning Simple Aperture Photometry (PDCSAP) flux for our analysis. PDCSAP light curves are generated light curves from SAP flux, which are removed of systematics effects using the Presearch Data Conditioning (PDC) pipeline module (see more details: \citealt{Jenkins2016SPIE.9913E..3EJ}). Additionally, we utilize the 'hardest' bitmask filter in \texttt{lightkurve} and then filtered the data by removing outliers and NaNs values and further normalized the light curves by dividing each flux by its mean.

Visual inspection of the PDCSAP light curve of MHO 4 in Figure \ref{fig: lc_ls_phs} displays a variable nature over the sectors. Figure \ref{fig: lc_ls_phs} also shows the periodogram and phase folded light curve of MHO~4.
Here, we utilized two independent techniques to infer the rotation period of MHO~4 using TESS data, i.e., Lomb-Scargle periodogram \citep{VanderPlas2018ApJS..236...16V} and Gaussian Process  {(GP)} \citep{Angus2018MNRAS.474.2094A}. We used \texttt{lightkurve.periodogram} to estimate the object's period, and the phase curve is constructed using the most significant peaks in the periodogram. We also attempted to infer the rotation period using the  {GP} method and provided a posterior probability distribution function that can be utilized to estimate rotation periods' uncertainty. Furthermore, for the GP method, we used \texttt{exoplanet} \citep{Foreman-Mackey2021exoplanet:joss} and \texttt{celerite2} \citep{Foreman-Mackey2018celerite2} package to model rotation periods in each Sectors individually. We elected to use \texttt{PyMc3} package for posterior distribution as it offers several general models. The estimated rotation periods of MHO~4 from TESS 2-min cadence data are described in Section~\ref{sec: result}. Furthermore, \texttt{Allesfitter} code is utilized to model the detected complex flare events of MHO~4 via Bayesian evidence. 


\section{Results and Discussions} \label{sec: result}
\begin{figure}
    \centering
    \includegraphics[width=0.49\linewidth]{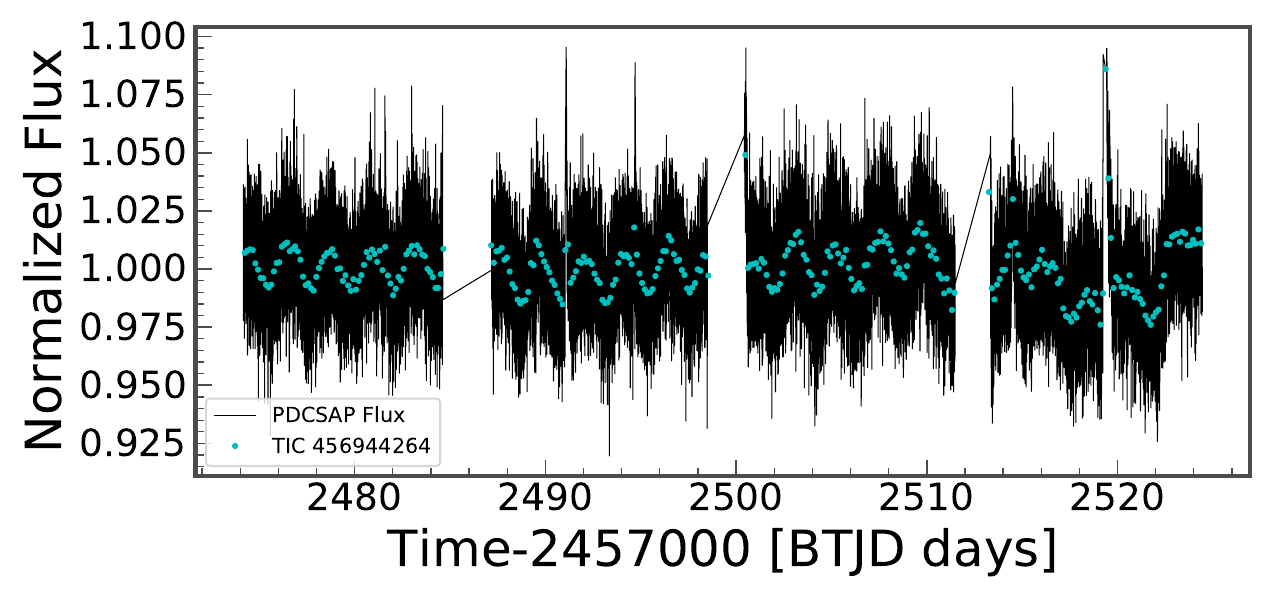}
    \includegraphics[width=0.47\linewidth]{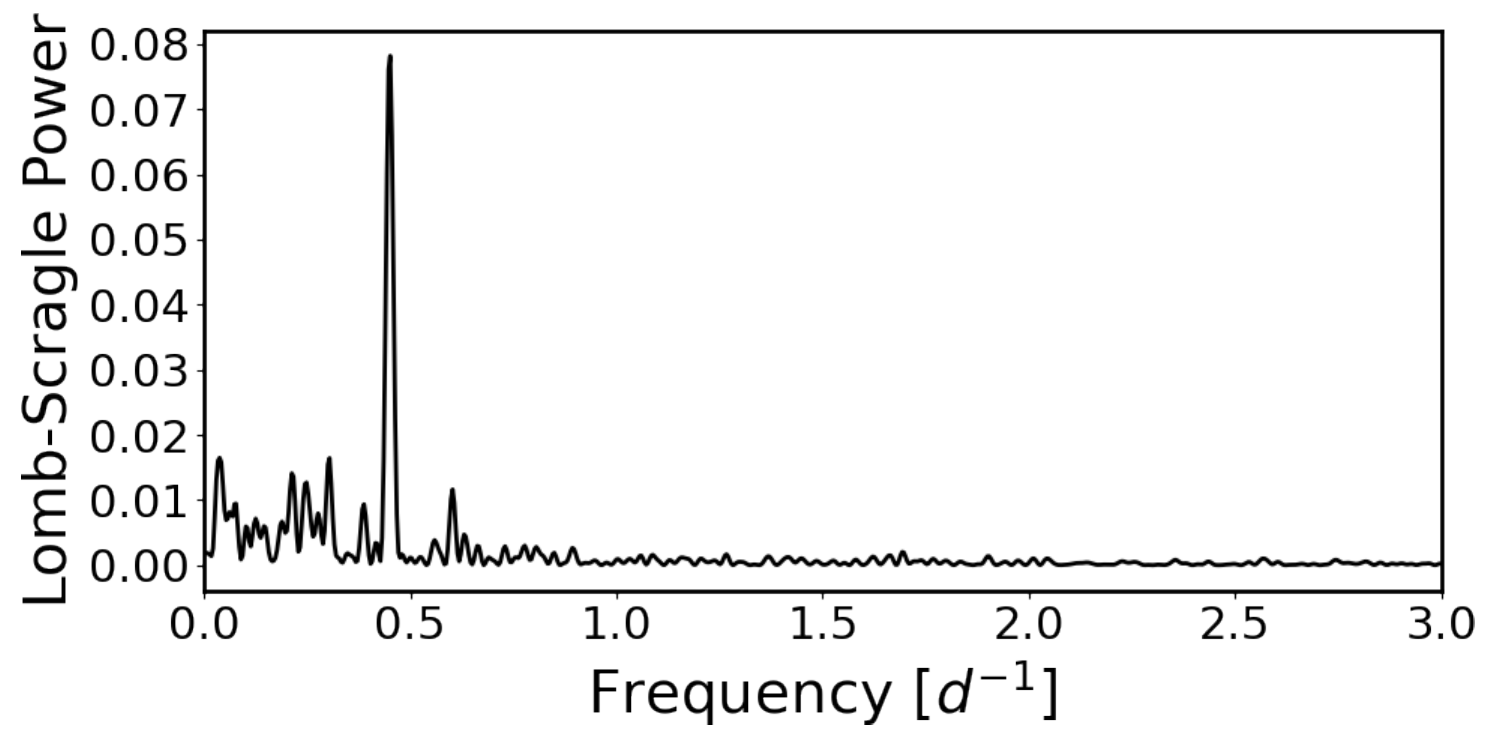}
    \includegraphics[width=0.50\linewidth]{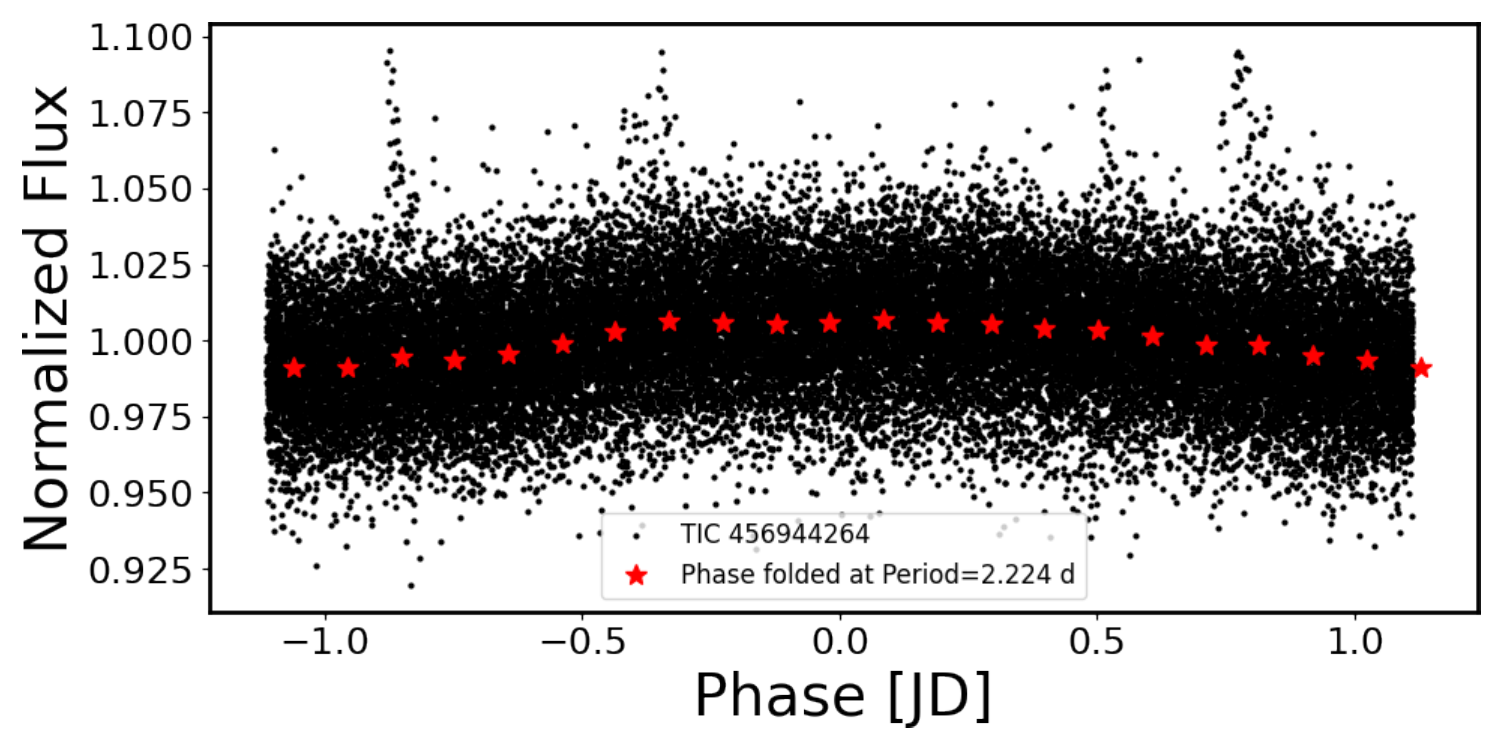}
    \caption{\footnotesize The TESS PDCSAP light curve  {(upper-left panel)}, LS periodogram  {(upper-right panel)} and phase folded light curve  {(lower panel)} of MHO~4 are shown here. Blue dots in the light curve represent the binning points of 200-min and red dots are the 150-min binning points in the phase curve. }
    \label{fig: lc_ls_phs}
\end{figure}

\begin{figure}
    \centering
    \includegraphics[width=6.5cm]{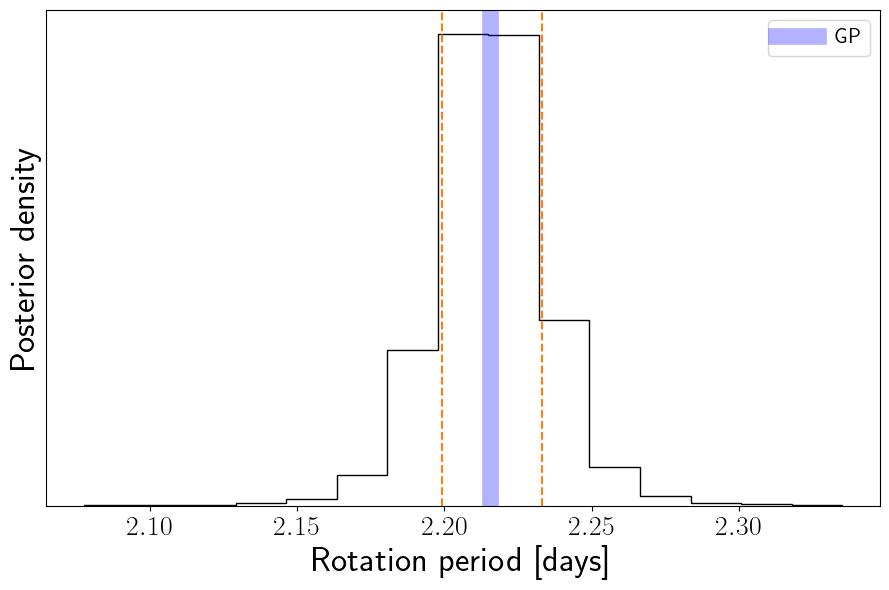}
    \includegraphics[width=6.5cm]{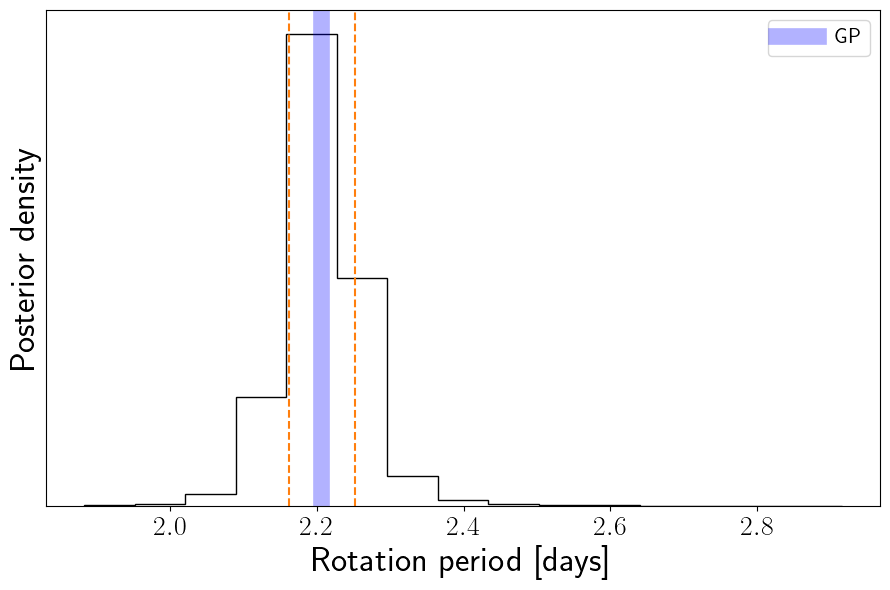}
    \caption{\footnotesize Posterior models of rotation periods of MHO~4, measured from TESS 2-min cadence data, is shown here. The blue line represents the rotation periods and the orange lines are the uncertainties to the periods.}
    \label{fig: gp_posterior}
\end{figure}

MHO~4 is a young BD having spectral type M7.0, $T_{eff}=2814$ K, $R=0.655 R_\odot$ \citep{Stassun2018AJ....156..102S}, belonging to the Taurus association. Previously, \citet{1998AJ....115.2074B} identified strong Li~I~$\lambda$6707 absorption line along with He~I~$\lambda$5876, [O~I]~$\lambda$6300, [O~I]~$\lambda$6363 for this object.
\citet{Rebull2020AJ....159..273R}, using the K2 observation, reported the period around 2.2098 d  while \citet{Gudel2007A&A...468..353G} recorded the upper limit of the rotation period as 6.29 d based on X-ray observation with the XMM-Newton data.
Figure \ref{fig: lc_ls_phs} illustrates the TESS light curve, LS periodogram, and phase curve of MHO~4 stitched to both sectors. 
We measured the rotation period using the LS method to be 2.224~d, and when analyzing each TESS sector separately, we also got consistent periods. The estimated rotation period using GP method is $2.215_{-0.016}^{+0.017}$~d in sector 43 and $2.205_{-0.043}^{+0.047}$~d in sector 44. The rotation periods derived from these two independent methods exhibit reasonably close values. The posterior distribution of rotation periods of MHO~4 for both sectors are shown in Figure \ref{fig: gp_posterior}. 

\subsection{Flare Detection and Analysis}
\begin{figure}[h]
    \centering
    \includegraphics[width=12cm]{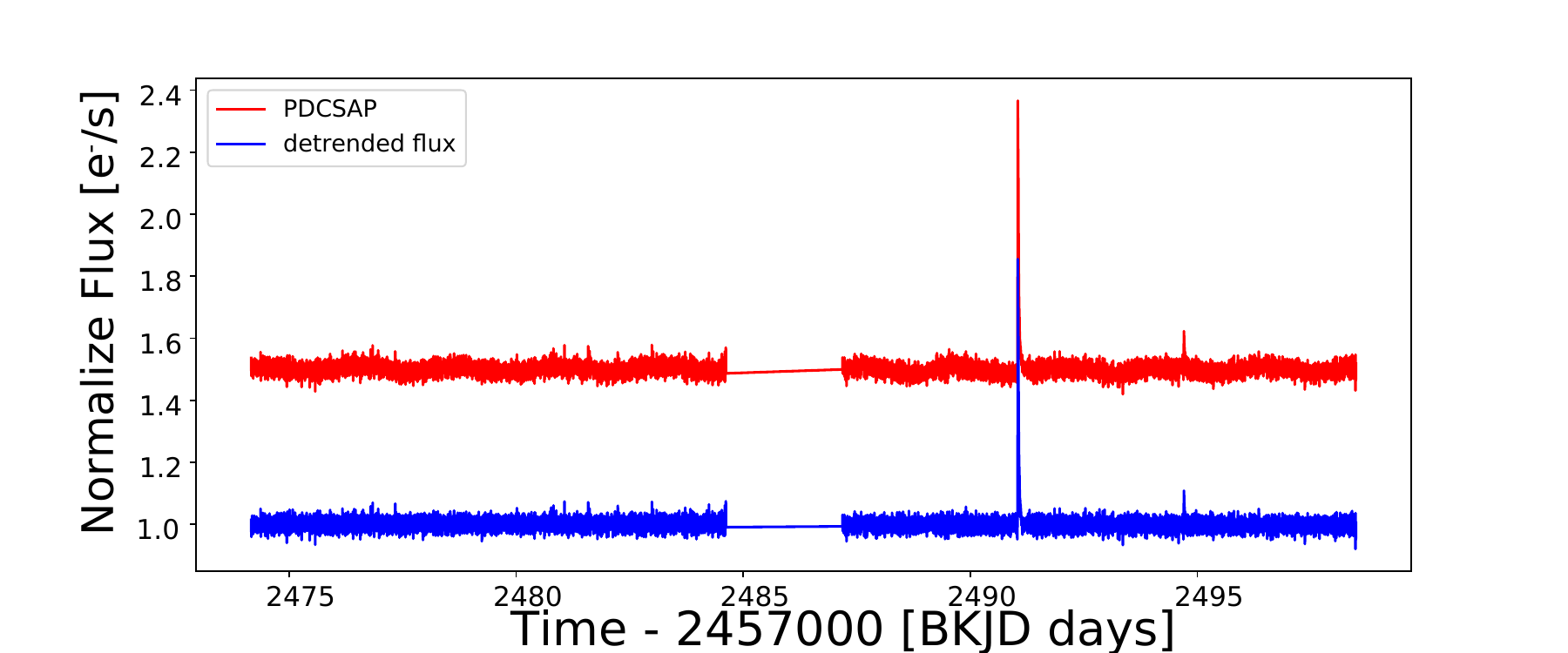}
    \caption{\footnotesize The PDCSAP and detrended light curves of MHO~4 for sector 44 are shown here. The x-axis is the time in Barycentric Kepler Julian Days (BKJD), and the y-axis is the normalized TESS flux ($\mathbf{e^{-} /s}$). Red highlights PDCSAP data. Blue gives the detrended flux ($\mathbf{e^{-} /s}$) with no intrinsic variation present..}
    \label{fig: detrn_lc}
\end{figure}

As a result of flare detection, we have identified one flares from each sector of MHO~4. For example, we presented the PDCSAP and detrended light curve for sector 44 in Figure \ref{fig: detrn_lc}.
Initially, we inspected visually and reconfirmed the detection of flares by an open source \textit{Python} software \texttt{ALTAIPONY} \citep{Ilin2021JOSS....6.2845I}. This software provides several flare parameters, like i.e., start and stop time, flare amplitude, and equivalent duration (ED). To estimate the bolometric flare energy, we used the equation defined by \citet{Shibayama2013ApJS..209....5S} and \citet{Ikuta2023ApJ...948...64I}. 
Assuming the flare as a blackbody of temperature 10,000 K, we calculated flare energy as $4.01\times 10^{34}$ erg in sector 43 and $ 1.59 \times 10^{35}$ erg in sector 44. Table \ref{tab: flare} describes the details of flare parameters.

\begin{table}
\centering
\caption{The details of flare properties of MHO~4}
\begin{tabular}{cccccccc}
\hline
Object & Sector & Radius & \begin{tabular}[c]{@{}c@{}}Temperature \\(K)\end{tabular} & Rel. Amplitude & ED (sec) & \begin{tabular}[c]{@{}c@{}}Energy\\  (erg)\end{tabular} & \begin{tabular}[c]{@{}c@{}}Duration\\    (min)\end{tabular} \\ \hline
MHO~4 & 43 & 0.655 & 2814.0 & 0.857 & $1479.79 \pm 11.10$ & 4.01E+34 & 84 \\
" & 44 & 0.655 & 2814.0 & 1.209 & $5852.15 \pm 22.17$ & 1.59E+35 & 298 \\ \hline
\end{tabular}
\begin{tablenotes}
  \item \footnotesize   {Note}: Radius and temperature are taken from \citet{Stassun2018AJ....156..102S} and amplitude (rel. flux) and EDs were computed using \texttt{Altaipony} software.
\end{tablenotes}
\label{tab: flare}
\end{table}

\begin{figure}[h]
    \centering
    \includegraphics[width=7.95cm]{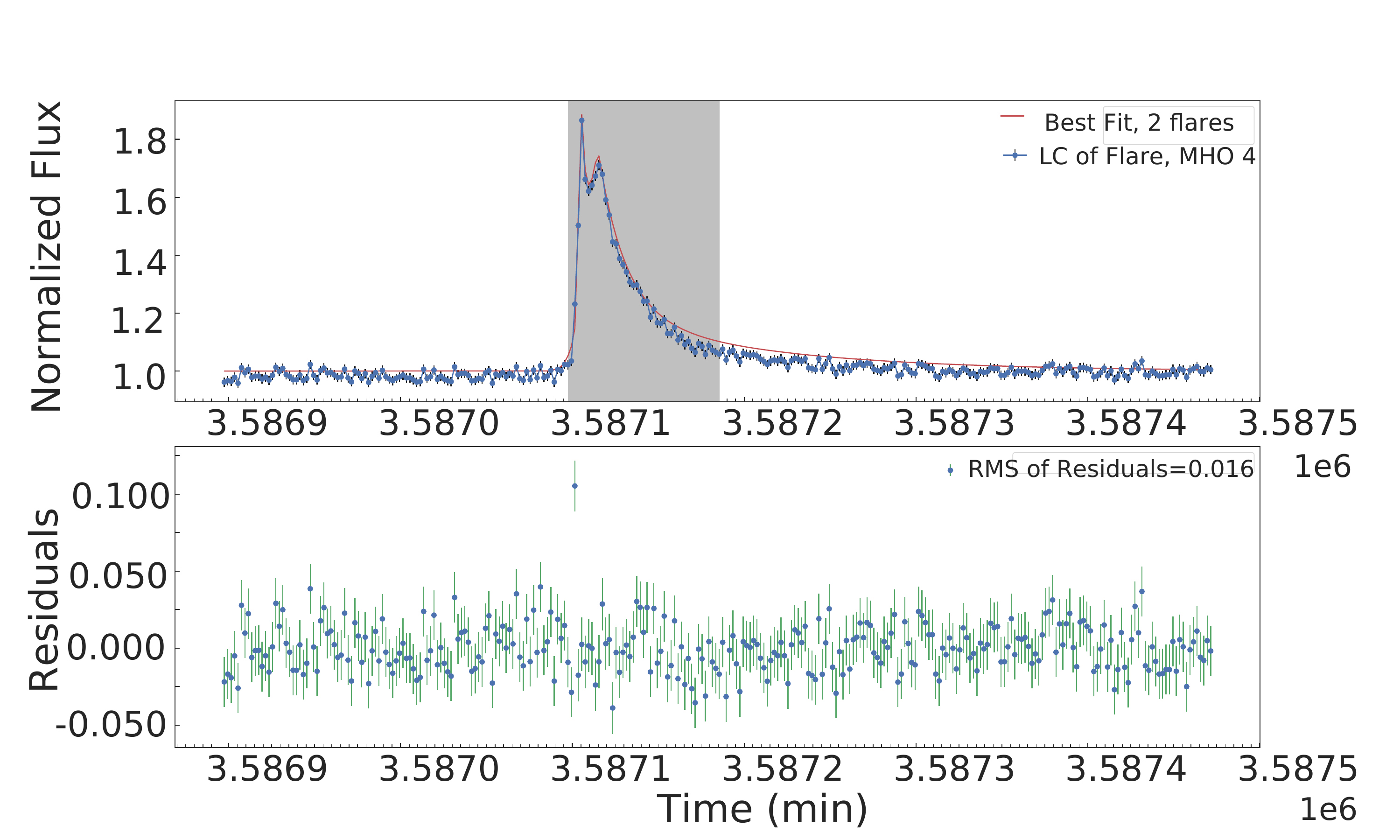}
    \includegraphics[width=7.95cm]{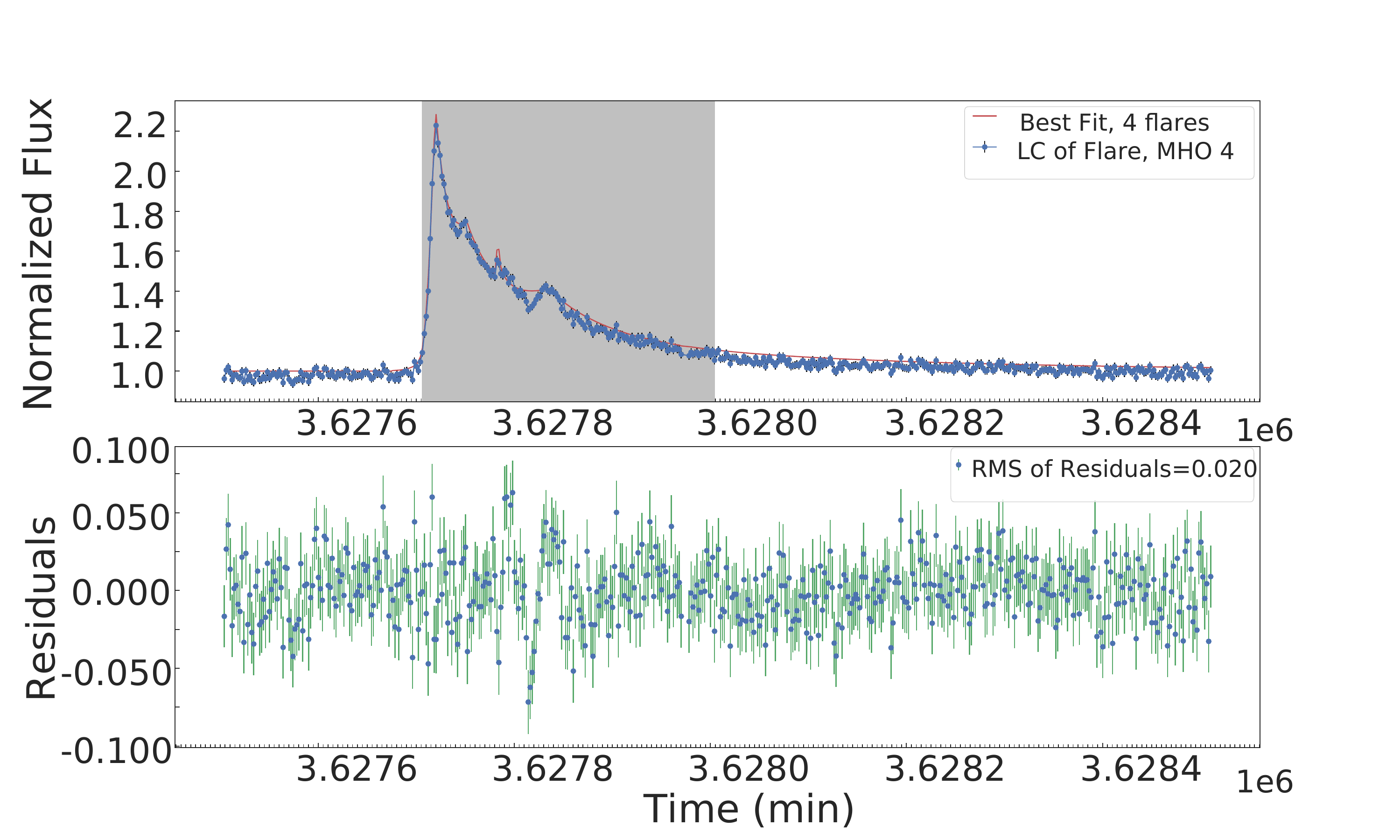}
    \caption{\footnotesize Top: The model fitted the flare light curve of MHO~4 in sector 43 (left) and sector 44 (right) using \texttt{Allesfitter}. Blue points are TESS 2-min observational cadence data, and the red line denotes the best-fitted curve. The x-axis shows the time in minutes y-axis shows the PDCSAP flux. Grey span indicates the region of the duration of flares. The below panel shows the residuals of the fitted light curves, and RMS values of the residual are shown in the legend.  }
    \label{fig: fitted_flc}
\end{figure}

\begin{figure}
    \centering
    \includegraphics[width=11.5cm]{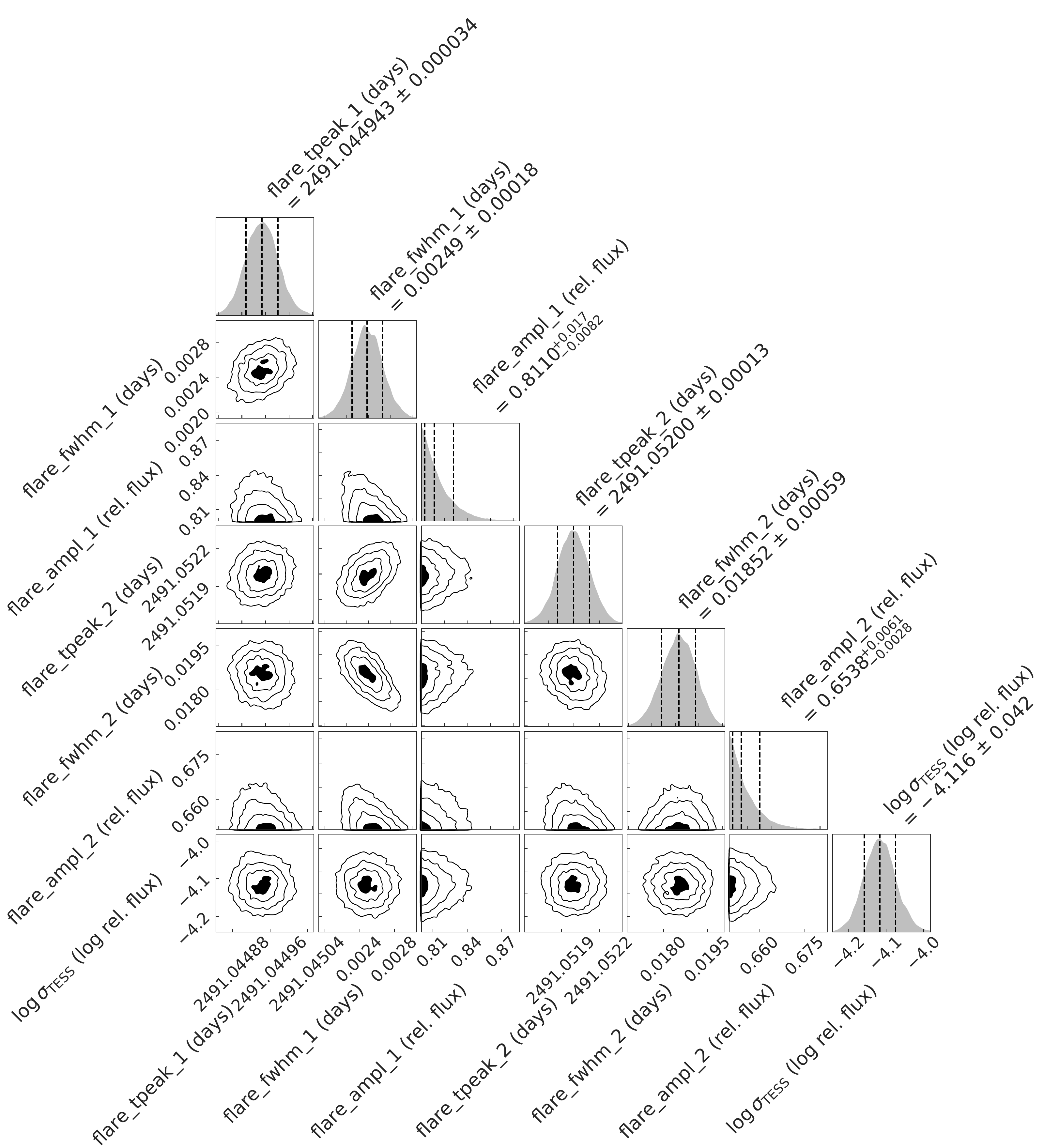}
    \includegraphics[width=11.5cm]{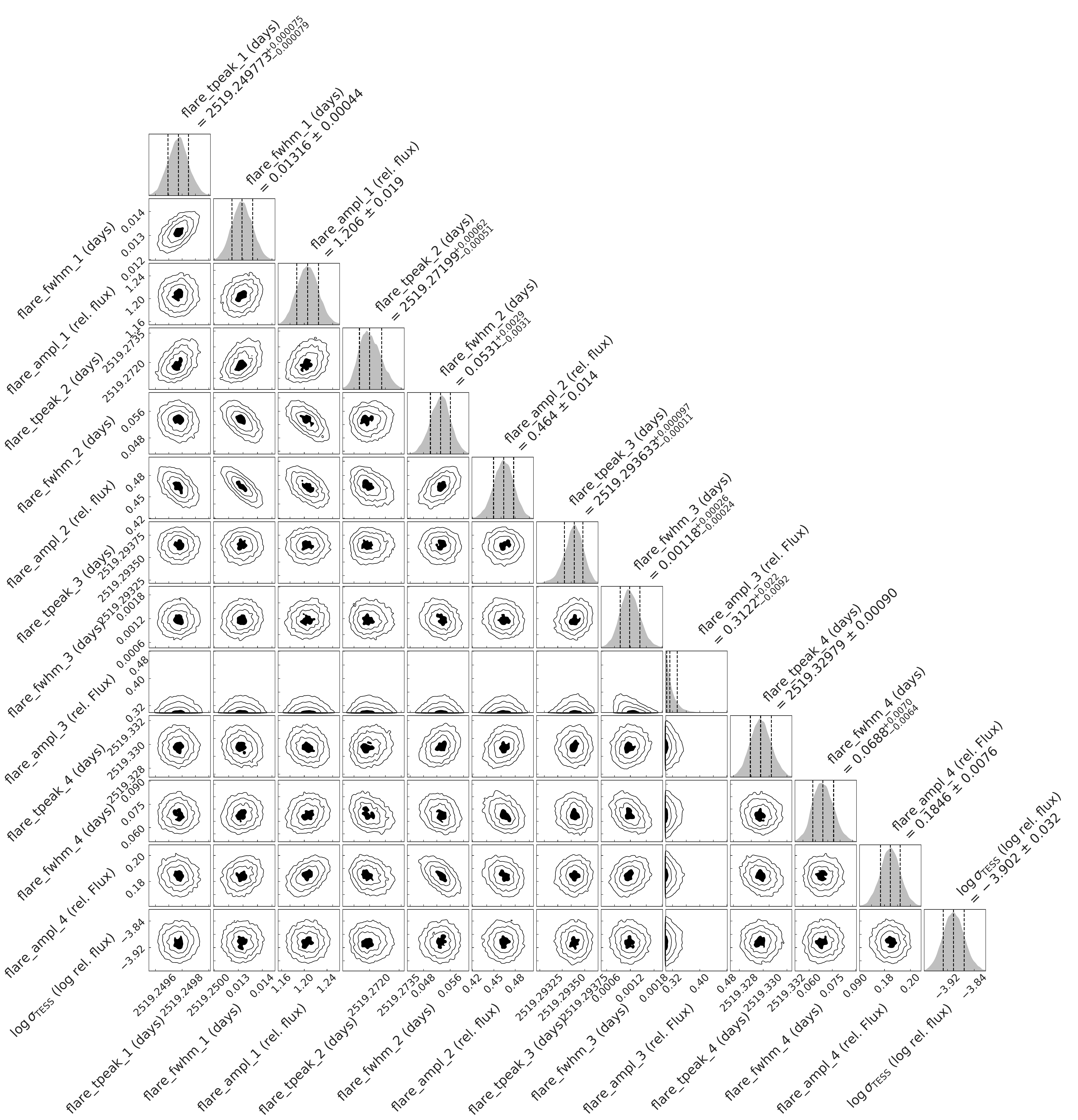}
    \caption{\footnotesize The posterior probability distribution of the flared light curve model parameters for sector 43 (left) and sector 44 (right).}
    \label{fig: posterior}
\end{figure}

Generally, flare light curves in BDs exhibit a rapid rise and gradual decline in brightness. However, all flare light curves show multi-peaked behaviour in the decline phase suggesting such flare events are complex. We further analyzed and modeled these flare light curves using open-source package \texttt{Allesfitter} (\citealp{Gunther2021ApJS..254...13G}) 
We modeled the flare light curve using the Nested Sampling \citep{Skilling2006article} algorithm, which provides the flares' peak times, amplitudes and FWHMs, white noise scaling, and a constant baseline. As a preliminary result, we found two flare candidate peaks in Sector 43, while in Sector 44, four flare candidate peaks fitted well to the flares. Figure \ref{fig: fitted_flc} illustrates the fitted flare light curve and their residuals. Figure \ref{fig: posterior} shows the posterior probability distribution of flare parameters.

Inspecting the power spectrum and phase light curve of MHO~4, we identified a significant peak in the LS periodogram. Folding the light curve at this period revealed a clear, smooth variation across the different sectors. Such smooth variations are typically associated with rotational modulation mainly of the spots or groups of spots with the objects \citep{Rebull2016AJ....152..113R}. So, here we deduced the mean spot temperature $T_{spot}$ of MHO~4 using a quadratic formula proposed by \citet{Herbst2021ApJ...907...89H}, $T_{spot}$ = - 3.58 $\times$ $10^{-5}$ $T^{2}_{eff}$ + 0.801$\times$ $T_{eff}$ + 666.5 

Additionally, We illustrated the spot relative intensity $f_{spot}$ for the TESS band, which depends on the stellar effective temperature $T_{eff}$ and the spot temperature,$T_{spot}$: $f_{spot}=\frac{\int R_{\lambda} B_{\lambda(T_{spot})} d\lambda}{\int R_{\lambda} B_{\lambda(T_{eff})} d\lambda}$

By applying these equations, we estimated the mean spot temperature of MHO 4 to be approximately 2638 K, with a spot relative intensity of approximately 0.65.

\section{Summary}
In this work, we investigated the photometric variability of a Taurus young BD, MHO~4, using the TESS 2-min cadence data to search for the rotation periods and atmospheric properties. 
From the analysis of the TESS data, we determined that MHO 4 has a rotation period of approximately 2.224 d, and its phase light curve exhibited smooth variations, which signified rotational modulation of asymmetrically distributed starspots/ groups of spots on its surface. We used two independent techniques, the Lomb-Scargle periodogram and the Gaussian process method, to infer the rotation periods,  {both yielding consistent results}. In addition, we detected two superflare events from MHO~4. The estimated bolometric energies of these flares were $4.01 \times 10^{34}$ erg (sector 43) and $1.59 \times 10^{35}$ erg (sector 44). Finally, to model the flare light curves, we observed that the flare in sector 43 was fitted with two flare candidate peaks, while four flare candidate peaks were needed to fit well in sector 44.

\begin{acknowledgments}
 We would like to thank Dr. Priya Hasan, the referee, for her valuable comments, suggestions on our paper which helped to improve the manuscript. This research work is supported by the S.N. Bose National Centre For Basic Sciences under the Department of Science and Technology, Govt. of India. This paper includes data collected with the TESS mission, obtained from the MAST data archive at the Space Telescope Science Institute (STScI). Funding for the TESS mission is provided by the NASA Explorer Program. STScI is operated by the Association of Universities for Research in Astronomy, Inc., under NASA contract NAS 5-26555. RK is grateful to the Department of Science and Technology (DST), Govt. of India, for their INSPIRE Fellowship scheme.
\end{acknowledgments}

\begin{furtherinformation}

\begin{orcids}
\orcid{0000-0001-7277-2577}{Rajib}{Kumbhakar}
\orcid{0000-0003-1457-0541}{Soumen}{Mondal}
\orcid{0000-0003-3354-850X}{Samrat}{Ghosh}


\end{orcids}

\begin{authorcontributions}
The light curves and power spectrum analysis were carried out by RK, SM and SG. The examination of the flare analysis part was performed by SG and RK. The text was written by RK. All of the authors provided input on the written draft of the manuscript as well as the discussion and interpretation of the findings.
\end{authorcontributions}

\begin{conflictsofinterest}
The authors declare no conflict of interest.
\end{conflictsofinterest}

\end{furtherinformation}

\bibliographystyle{bullsrsl-en}

\bibliography{extra}

\end{document}